\documentclass[12pt,preprint]{aastex}
\usepackage{natbib}
\def\ha{{H$\alpha$}}
\def\arcsec{$^{\prime\prime}$}

\begin{document}

\title{Explosive events associated with a surge}

\author{M.S. Madjarska$^{1 , 2}$, J.G. Doyle$^2$ \& B. de Pontieu$^3$}

\shortauthors{Madjarska et al.}
\shorttitle{Explosive events  associated with a surge}

\affil{$^1M$ax-Planck-Institut f\"ur Sonnensystemforschung, Max-Planck-Str. 2, 37191
 Katlenburg-Lindau, Germany}
\affil{$^2$Armagh Observatory, College Hill, Armagh BT61 9DG, N. Ireland}
\affil{$^3$Lockheed Martin Solar and Astrophysics Laboratory, 3251 Hanover Street, 
Organization ADBS, Building 252, Palo Alto, CA 94304, USA}
\email{madj@arm.ac.uk}

\begin{abstract}

The solar atmosphere contains a wide variety of small-scale transient features.
Here, we explore the inter-relation between some of them such as surges, explosive 
events and blinkers via simultaneous spectral  and imaging data taken with
the TRACE imager, the SUMER,  and  CDS spectrometers on board SoHO, 
and SVST La Palma. The features were observed 
in spectral lines with formation temperatures from 10,000~K to 1~MK and with 
the TRACE Fe~{\sc ix/x}~171~\AA\ filter. The \ha\ filtergrams were taken in the wings of  the
\ha\ 6365~\AA\  line at $\pm$700~m\AA\ and $\pm$350~m\AA. The alignment of all
 data both in time and solar XY shows that SUMER line profiles, which are attributed
to  explosive events,  are due to a surge phenomenon. The surge's up- and down-flows 
 which often appear simultaneously correspond to the blue- and red-shifted emission 
 of the transition region N~{\sc v}~1238.82~\AA\ and O~{\sc v}~629.77~\AA\  lines as 
 well as radiance increases of the C~{\sc i}, S~{\sc i} and S~{\sc ii} and Si~{\sc ii} 
 chromospheric lines. Some parts of the surge are also visible in the TRACE~171~\AA\ 
 images which could suggest heating to coronal temperatures. 
 The surge is triggered, most probably,  by one or more  Elerman bombs which are best visible 
 in \ha\ $\pm$350~\AA\ but were also registered by TRACE Fe~{\sc ix/x}~171~\AA\ and 
 correspond to a strong radiance increase in the
  CDS Mg~{\sc ix}~368.07~\AA\ line.  With the present study we demonstrate that  the division of 
small-scale transient events  into a number of different subgroups, for instance explosive 
events, blinkers, spicules, surges or just brightenings, is ambiguous, implying that the 
definition of a feature based only on either spectroscopic or imaging characteristics  
as well as insufficient spectral and spatial resolution can be incomplete.

\end{abstract}


\section{Introduction}

 Over the last decades, the increased use of space observatories such as the 
Solar and Heliospheric Observatory (SoHO), the Transition Region Coronal Explorer 
(TRACE) and presently Hinode, coupled with ground-based observations, have brought many 
new features on the Sun to light. What is, however, more important is that
the combination of different types of observations, i.e. spectroscopic and imaging,
covering a wide wavelength range and having a similar cadence and spatial resolution,  
provide crucial information on the nature and the physical characteristics of these 
phenomena.

We now know that the solar atmosphere contains different kinds of small-scale transient 
events. A particular transient phenomenon, however, is often associated with a specific 
instrument or a type of instrument, either a spectrometer or an imager. Explosive events 
(EEs; registered by the Solar Ultraviolet Measurement of Emitted Radiation (SUMER) 
spectrometer  and the High Resolution Telescope and Spectrometer (HRTS)), for instance, 
were found  to be restricted to transition region temperatures from  
2~$\times$~10$^{4}$~K to 5~$\times$~10$^{5}$~K. They  were first discovered by Brueckner \& Bartoe (1983) in 
C~{\sc iv}~1548.21~\AA\ data taken with HRTS. They are characterized as short-lived 
(60 -- 350~{\rm s}) small-scale (3\arcsec\ -- 5\arcsec\ along a spectrometer slit) 
events identified by Doppler shifts of up to  200~km~s$^{-1}$.  It has been suggested 
that they  result from the production of high-velocity bi-directional plasma jets 
during magnetic reconnection \citep{dere}. They often occur in areas with weak fluxes 
of mixed polarity or on the border of regions with large concentration of magnetic flux 
\citep{chae1998} and are often observed in bursts lasting up to 30~{\rm min}
in regions undergoing magnetic cancellation (Dere 1994, Chae et al. 1998, Perez et al. 1999, 
Doyle et al. 2006). Another term often associated with them is bi-directional jets 
\citep{innes}. 

Surges, mainly observed in \ha\ (10$^4$~K) and  Ca~{\sc ii}~K \& H (6 10$^3$~K) as well as 
Extreme-ultraviolet (EUV) and  X-ray jets are also assumed to result from the same physical
 mechanism. Solar surges  are associated with active regions and represent an ejection 
 of dense chromospheric  material (N$_e$ $\approx$ 10$^{11}$~{\rm cm}$^{-3}$) in the shape of straight or 
arch-shaped collimated  streamers seen dark or bright in the \ha\ line \citep{roy}. 
The up-flows  often return to the solar surface along the same trajectory. Their 
occurrence is associated with Ellerman bombs \citep{ellerman} in their foot-points.
 
The term blinker  was first used by Harrison (1997) to describe a small-scale 
(on average 8\arcsec\  $\times$ 8\arcsec)  brightening in EUV lines observed with the  
Coronal Diagnostics Spectrometer (CDS) onboard SoHO. They last on average 17~{\rm min}  
and are preferentially located in the network boundaries \citep{bewsher2002}.  
It has been suggested that they represent an observational signature of increased 
filling factor rather than electron density. Several physical mechanisms have been 
put forward by Priest et al. (2002), Doyle et al. (2004) and Marik \& Erd{\'e}lyi (2002).

There have been attempts to unify some of these terminologies, e.g. blinkers have
been suggested as a generic term for EUV network and cell brightenings 
\citep{harr2003}. Chae et al. (2000) stated that blinkers 
and EEs were the same  phenomenon, while Madjarska \& Doyle (2003) found them 
to be two separate unrelated events, as did a statistical study by 
Brkovi\'c  \& Peter (2004).  Madjarska \& Doyle (2003)  also speculated that 
blinkers may simply be the on-disk signature of spicules. Their work later lead 
to the suggested notion that  some blinkers and macro-spicules are the same 
phenomenon \citep{oshea, madj2006}.

The uniqueness of the present study lies in the exploration of small-scale transient 
phenomena using simultaneous EUV high-cadence and the highest existing spatial and 
spectral resolution data currently available together with ground-based \ha\ observations.
We analyzed numerous small-scale transient features seen along 
a SUMER slit in chromospheric (C~{\sc i}, Si~{\sc i}, {\sc ii} and S~{\sc ii}), 
transition region (N~{\sc v} and O~{\sc v}) and coronal (Mg~{\sc x}) spectral lines, 
and their counterparts  in SVST \ha\ filtergrams and TRACE 171~\AA\ images. Our 
main objective is to derive the spectral characteristics of these events and compare 
them with their image appearance at  coronal and chromospheric temperatures which 
should provide their correct identification and evaluation of their  plasma quantities and 
dynamics, and could suggest a possible physical mechanism of their generation.

\section{SUMER, CDS, TRACE and SVST La Palma observational material}

All data were obtained on 1 June 1999 in active region NOAA~8559  with the SUMER 
slit placed over the plage area. The standard  data reduction procedures were  
applied to all observational material. The SUMER dataset \citep{wilh}  was 
taken  from 09:03:23~UT to 11:01:42 UT~with the solar rotation compensation mode turned 
on.  Slit  0.3\arcsec\ $\times$~120\arcsec\ was used, exposing for 25~{\rm s} the bottom part of  
detector B. A list of the analyzed spectral lines is given in Table~\ref{table1}.  The CDS  
data  were taken with a 25 {\rm s} exposure time resulting in a 34~{\rm s} cadence in a sit-and-stare 
mode with a 2\arcsec $\times$ 181\arcsec\ slit.  The observed spectral lines are also reported in 
Table~\ref{table1}. The TRACE~171~\AA\  dataset \citep{handy} was obtained from 09:15:44~UT to 
10:39:56~UT, exposing for  5.8~s resulting in 10~s to 40~{\rm s} cadence. The images were derotated 
to a reference time of 08:17~UT.  The \ha\ 6563~\AA\  filtergrams were made with the 
Swedish Vacuum Solar Telescope (SVST) in La Palma. The \ha\ data were registered at $\pm$350 m\AA\ 
and $\pm$700 m\AA\ and  have a pixel size of 0.083 arcsec/px. The cadence of the \ha--350~m\AA\ and --700~m\AA\  data
is 1m14s$\pm$1s and  of the \ha+350~m\AA\ and +700~m\AA\ data  1m11s$\pm$2s.  In Fig.~\ref{fig1} we show 
an overview of the region observed by TRACE  and SVST 
 with the locations of the SUMER and CDS slits.  The alignment of SUMER, CDS to the TRACE data with a 
 precision of $\pm$1\arcsec\ was done as described in Doyle et al. (2005), using TRACE 171~\AA\ images and a 
CDS Mg~{\sc ix}~368~\AA\ raster. The coalignment of the \ha\ data to TRACE was done 
according to the recipe in de Pontieu et al. (1999).  

\section{Results \& Discussion}

First, we inspected the SUMER spectral line profiles looking for radiance increases  
and/or line broadenings. Fig.~\ref{fig2} shows the radiance variations of the transition  
region O~{\sc v}~629~\AA\ and N~{\sc v}~1238~\AA\ lines. Brightenings are clearly seen 
in the area indicated with the horizontal dashed lines. The line profiles reveal 
a different response of the two transition region lines O~{\sc v} and N~{\sc v} which 
will be discussed later in the paper. The images were produced from the peak radiance 
of the spectral lines obtained from a single Gauss fit with the continuum emission 
removed. That helps to separate the intensity variation in the spectral line from the 
continuum emission changes. Increases of the continuum emission during explosive events 
is a well know fact \citep{madj2002}.
 
The next step of our analysis was to produce animated time sequences from some of the 
available data (see the on-line material). The movie consists of TRACE~171~\AA\ images, 
the SUMER spectral windows of O~{\sc v} 629~\AA, Mg~{\sc x}~625~\AA\ together with 
several chromospheric lines, plus N~{\sc v}~1238~\AA, and SVST \ha\ 6365~\AA\ 
filtergrams taken at --350~m\AA. Additional movies can be seen at URL 
\url{http://www.arm.ac.uk/$\sim$madj/outgoing/Surge}. 

\begin{figure}[hbt!]
\epsscale{0.8}
\includegraphics{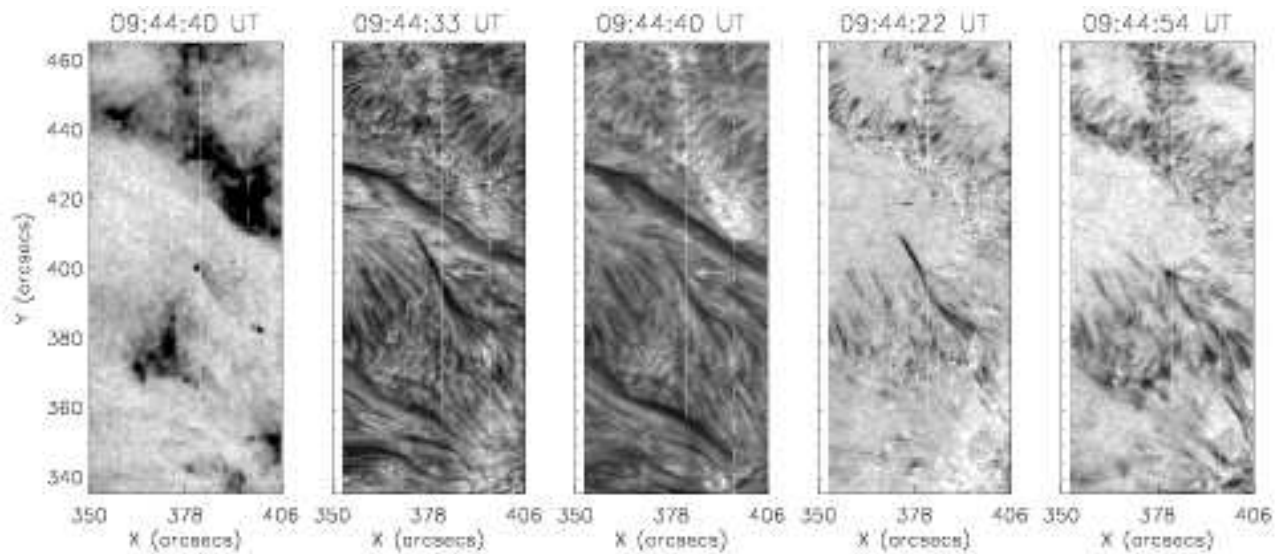}
\vspace{-2.5cm}
     \caption{Overview of the observed region taken (from left to right) with 
     the TRACE~171~\AA\ filter, \ha\ --350, +350, --700 and +700~m\AA. The vertical 
     solid line indicates the SUMER slit location, while the vertical dashed the CDS 
     slit position. The arrow indicates the elongated structures  which were identified as a surge.
       We should note that  although SUMER and CDS were commanded to 
     point at the same solar disk co-ordinates, the two slits are actually offset by 14\arcsec -- 15\arcsec. 
     One possible explanation is that SUMER had lost motor steps during the pointing procedure (W. 
     Curdt, private communication). The TRACE image is shown with reversed color table, i.e.  
     darker structures mean stronger emission.}
   \label{fig1}
\end{figure}

\begin{figure}[hbt!]
 \epsscale{1.1}
\plottwo{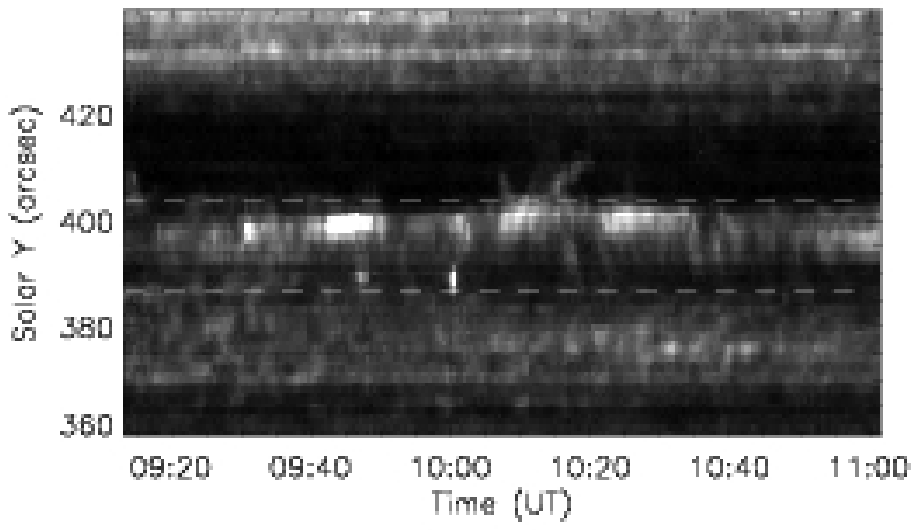}{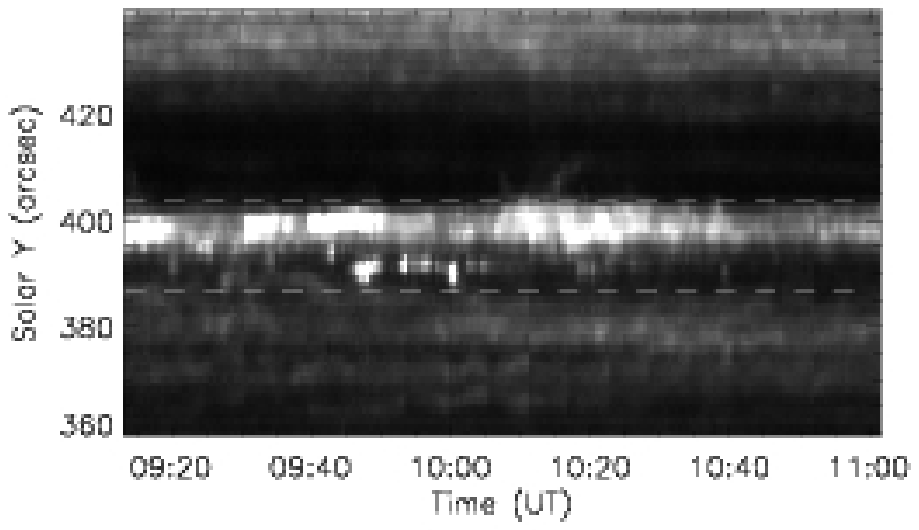}
\caption{The radiance variations of the O~{\sc v} (left) and N~{\sc v} (right) lines along the 
SUMER slit. The horizontal lines indicate the area used to analyze the spectral line 
profile variations. The light curves shown in  Fig.~\ref{fig3} and \ref {fig4} were produced from this region.}
\label{fig2}
\end{figure}

\begin{figure}[hbt!]
\includegraphics{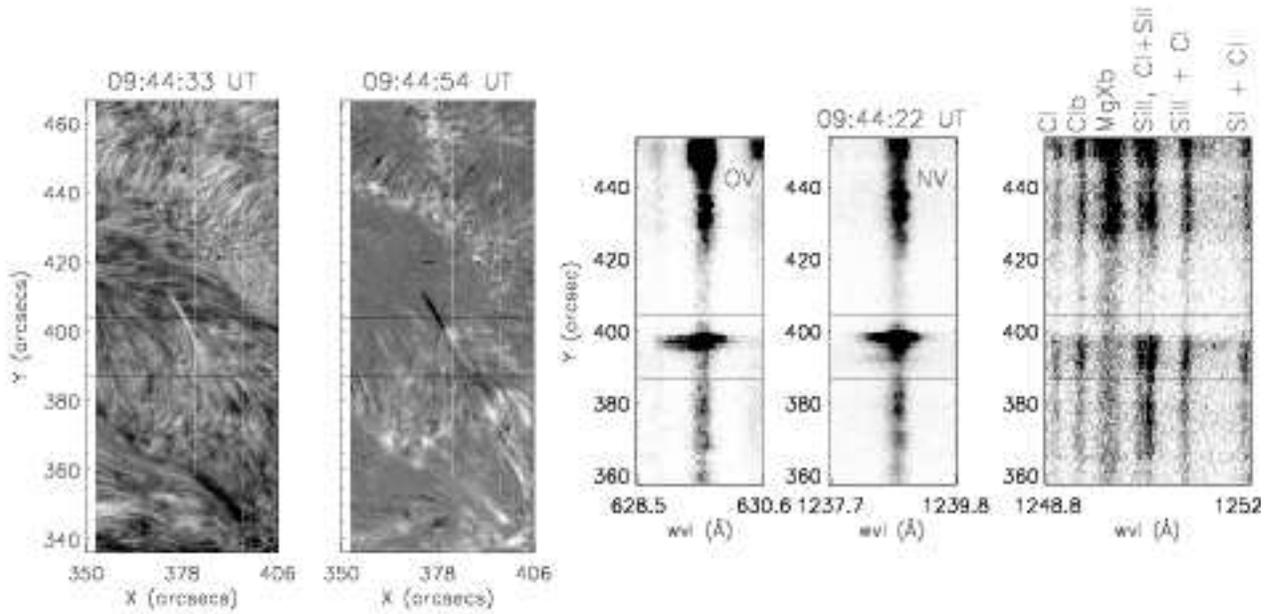}
\caption{Doppler images from \ha\ 6365 \AA\ $\pm$350~m\AA\ (first from left) and 
$\pm$700~m\AA\ (second), followed by O~{\sc v}~629~\AA, N~{\sc v}~1238~\AA\ and the 
spectral window covering Mg~{\sc x}~625~\AA\ and several chromospheric lines, as indicated on the image (see also Table~1.)
The vertical solid line indicates the 
SUMER slit location, while the dashed vertical line the CDS slit position. The horizontal solid line indicates the region 
from which the SUMER light curves were produced. The arrow points at the surge.
     \label{fig_new_add}}
\end{figure}

The SUMER spectral line profiles during the brightening events show all 
the hallmarks of explosive events, i.e. non-Gaussian profiles and  only blue- or 
red-shifted emission or both.  An example of these profiles is shown in Fig~\ref{fig_new_add}.
 When coupled with the TRACE~171~\AA\ images, we see that 
some of the non-Gaussian profiles correspond to flows along elongated structures (see {Fig.~\ref{fig1}}). 
However, not all SUMER line brightenings or wing emission increases correspond to an 
observed feature in the TRACE 171~\AA\  images. The simple explanation is that not all 
structures reach temperatures corresponding to the temperature response of the 171~\AA\ 
filter. The simultaneously obtained \ha\ images reveal that the \ha\ counterpart of 
the TRACE and SUMER events are numerous up- and down-flows, often happening simultaneous,  
seen in images taken at $\pm$350 and $\pm$700~m\AA.  Additionally, we 
constructed  Doppler images using the relation  $Dopp = (C_b-C_r)/(C_b+C_r+2)$ \citep{suem95}
where ${C_b}$ and ${C_r}$ are the contrast images obtained as $C = (I - I_a)/I_a$ . The $I_a$  corresponds to 
the average intensity over the whole dataset. The subscripts {\rm b} and {\rm r} correspond to   the wavelengths 
at \ha\ 6563~\AA\ --350~m\AA\ (or --700~m\AA) and +350~m\AA\ (or +700 m\AA), respectively (see Fig~\ref{fig_new_add}).  
The dark emission from the surge corresponds to the \ha\ +350 m\AA\ and +700 m\AA\ while the bright to \ha\  -350 and -700.  
With the available information from the \ha\ line it is not possible to derive  (for instance via cloud model, see 
Alissandrakis et al. (1990) and Tsiropoula \& Schmieder  (1997))  the full meaning of the bright and dark features.

\begin{figure}
\epsscale{1.0}
\plotone{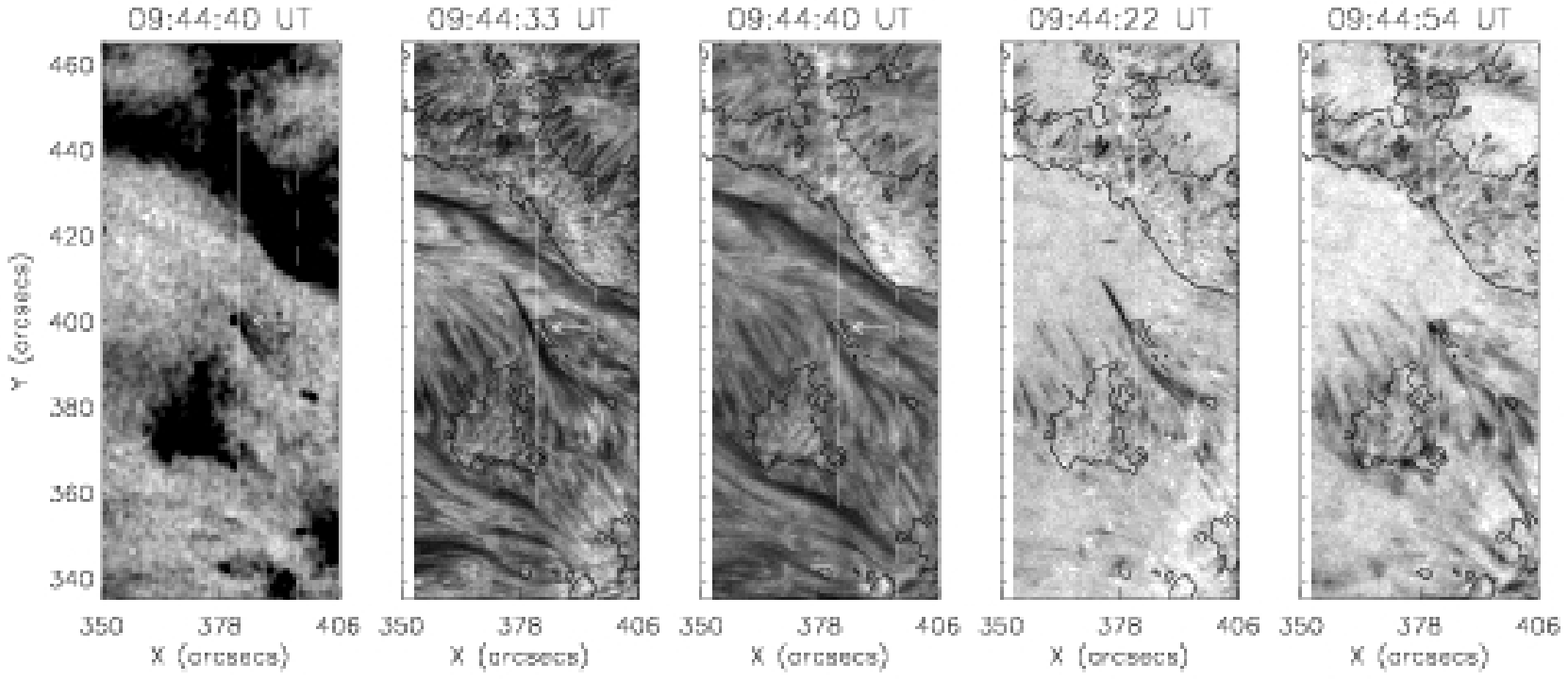}
\plotone{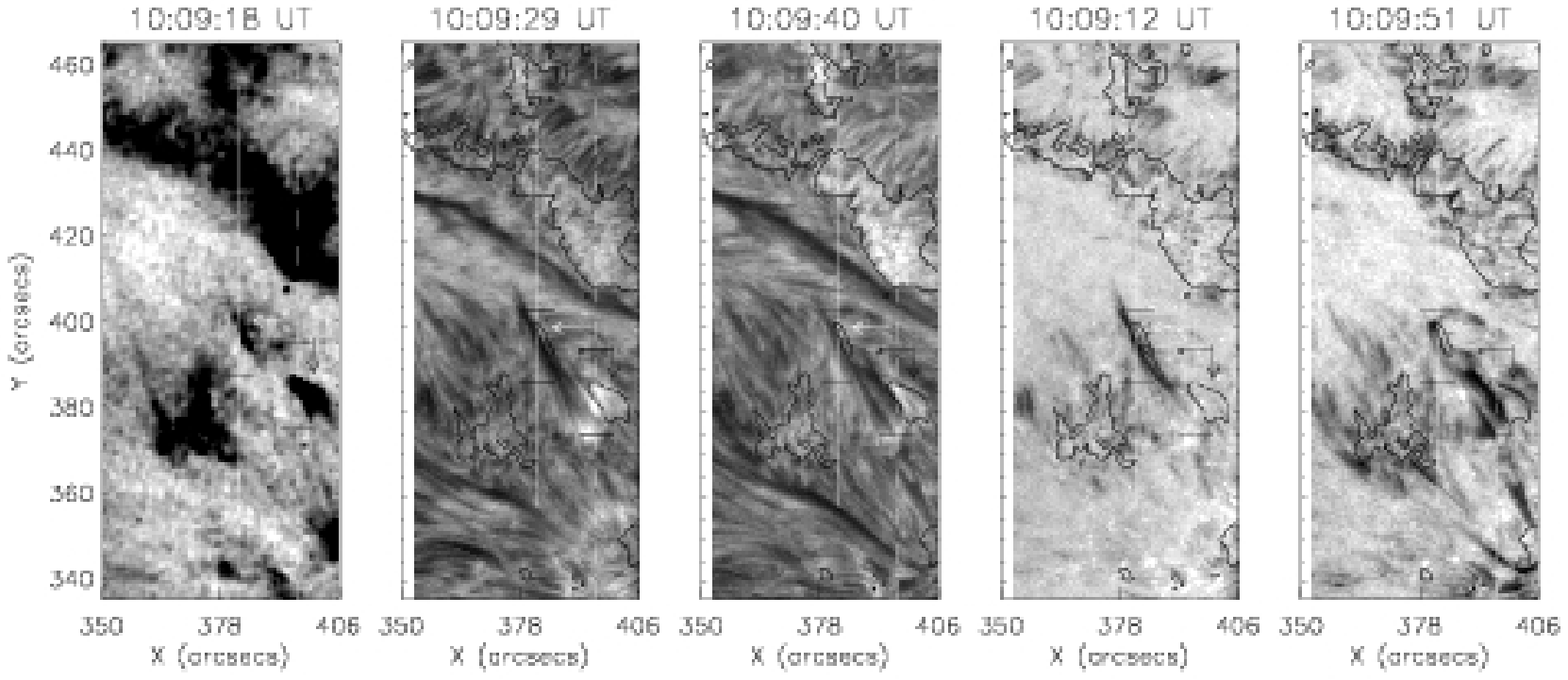}
\caption{TRACE~171~\AA\ images and \ha\ --350, +350, --700, +700~m\AA\  
filtergrams with over-plotted the TRACE contours  taken at two 
different moments in time during the surge.  The white horizontal arrow points at the 
upper part of the surge. The black vertical arrow indicates the Ellerman 
bomb (only  at around 10:09~ UT). The vertical solid line indicates the 
SUMER slit location, while the dashed vertical the CDS slit position.}
\label{fig6}
\end{figure}

These up- and down-flows were identified as 
a typical surge event. The foot-points of the surge are anchored in one or more  Ellerman bombs
which re-occur several times triggering the surge re-appearance,  as shown in Fig.~\ref{fig6}, where  we present a set of 
all images (TRACE plus all four \ha\ images) taken at two different times. The TRACE contours
over-plotted on the \ha\ images help to cross-correlate  the feature as seen in TRACE with the dark and bright \ha\ 
features from the surge and its foot-points.   Jiang et al. (2007) 
presented in great detail the relation between \ha\ surges, EUV and X-ray jets which 
resemble very much the event analysed here (Fig.~\ref{fig6}). They concluded that each 
surge consisted of a cool \ha\ component and hot, EUV or soft X-ray component which showed 
different evolution in space and time. The cadence of our \ha\ data does not permit such 
an analysis. However, the SUMER data do not show evidence for a different spatial  behavior 
of the chromospheric and transition region emission. We should note that this could be due to
 the difference in the spatial resolution between the SVST and SUMER data
 as well as the fact that slit spectrometer data have their limitation for space analysis. 
 Not all features seen in the transition region
lines, however,  have their counterpart in the chromospheric lines which suggests plasmas heated to 
higher temperatures.  Unfortunately,
the lack of magnetic field data makes it impossible to discuss  the photospheric 
magnetic field configuration. 

\begin{figure}
\epsscale{0.6}
\plotone{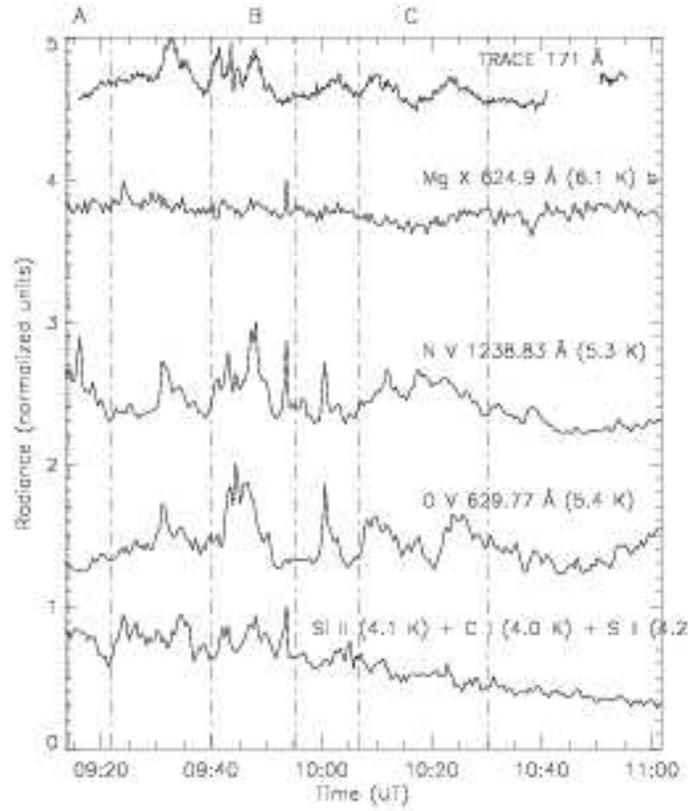}
\caption{Normalized light curves for Si~{\sc ii} 1250.40~\AA\ + C~{\sc i}  1250.42~\AA\ + 
S~{\sc ii} 1250.58  (the first from the bottom), O~{\sc v} 629.77~\AA, N~{\sc v} 1238.82~\AA,  
Mg~{\sc x}~624.9~\AA\ and TRACE Fe~{\sc ix/x}~171~\AA. The formation temperature in $Log\,T$ 
of each spectral line is shown in brackets. The mark ``b'' means blended.}
\label{fig3}
\end{figure}

We performed a further spectroscopic analysis in order to gather more information on 
the relation between explosive events and surges. In order to analyze the temperature 
response of the surge plasma, we produced normalized light curves (Fig.~\ref{fig3}) 
from spectral lines covering a large range of formation temperatures  (see Table~\ref{table1}). 
The light curves were obtained by binning the partition along the SUMER slit outlined by 
two horizontal lines in Fig.~\ref{fig2} corresponding to 18 pixels ($\sim$18 arcsec).  The TRACE light 
curve was obtained by binning over an array of 4\arcsec$\times$4\arcsec\ at Solar X = 382\arcsec\ 
and Solar Y = 400\arcsec\  where the SUMER slit crosses the event(s) as indicated in Fig.~\ref{fig1}. 
All light curves were normalized to the maximum value of the radiance during the studied 
time interval. The TRACE 171~\AA\ pass-band, although dominated by the Fe~{\sc ix} and Fe~{\sc x} lines, 
has some transition region lines in the range of 171.5--174~\AA, e.g. O~{\sc v} and O~{\sc vi}, though, as shown 
by Doyle et al. (2003), their effect is small unless there is a substantial 
departure from a Maxwellian velocity distribution. Among the observed spectral lines 
is the coronal Mg~{\sc x} 625~\AA\ (1~ MK) line. Its light curve in Fig.~\ref{fig3} shows no 
clear evidence for enhanced emission except around 09:25~UT and
09:55~UT. Comparing with the chromospheric light curve in Fig.~\ref{fig3}, we found
that these increases in the Mg~{\sc x} emission are due to the 
blend from several cooler lines (see Table~\ref{table1}) rather than Mg~{\sc x}
itself  \citep{teriaca}.  Another possible explanation is that Mg~{x}  
has very slow ionization and recombination time scales, as described by 
Golub et al. (1989).  We cannot speculate on the  decrease in the light curve of  the chromospheric lines 
for the time period 10:30 -- 11:00~UT.

Chae et al. (1998), using SUMER data in Si~{\sc iv} 1402~\AA\ and Big Bear Solar 
Observatory (BBSO) \ha\ spectrograph observations, found that chromospheric upflow events 
arising in intranetwork areas are related to transition region explosive events.  
Madjarska \& Doyle (2002) studying high-cadence (10~{\rm  s}) data 
during explosive events in the chromospheric Ly~6 960.75~\AA\ line (20\,000~K) and 
the transition region  S~{\sc vi}~933.38~\AA\ (200\,000~K) line found a time delay in the 
response of the S~{\sc vi} line with respect to the Ly~6 line, with Ly~6 responding 
20-40~{\rm s} earlier. That made the authors suggest heating from chromospheric to 
transition region temperatures, i.e. a chromospheric origin of explosive events. The 
complexity of the event(s) analyzed in the present  work makes it impossible to 
temporally resolve the connection of cold and hot plasmas. It is also possible that 
the time resolution of the present data is not sufficient  for such analysis.  

\begin{figure}
\epsscale{0.6}
\plotone{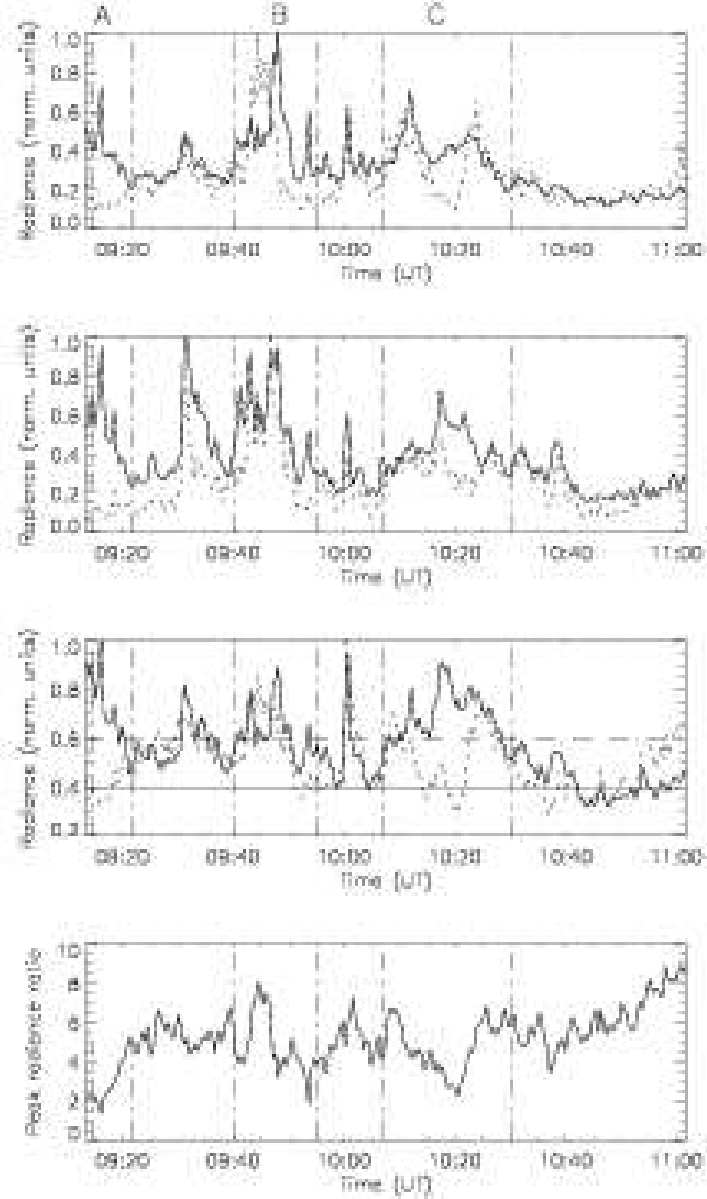}
\caption{Normalized light curves in the blue wing (first from top), red wing
(second) and 
the core of N~{\sc v} 1238.82~\AA\ (solid line) and O~{\sc v} 629.77~\AA\ (dotted line). 
The solid horizontal line in the third panel represents the background emission while 
the dashed horizontal line, 1.6 times the background emission.The bottom panel presents the radiance ratio
 of O~{\sc v}/N~{\sc v}.}
\label{fig4}
\end{figure}

\begin{figure}
\epsscale{0.6}
\plotone{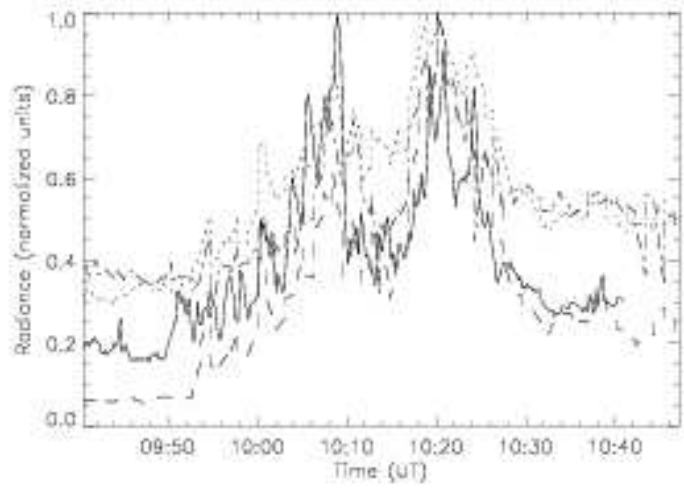}
\caption{Normalized light curves of the peak radiance  of TRACE~171~\AA\ (solid line),  
CDS O~{\sc v} 629.77~\AA\ (dashed line), CDS Mg~{\sc ix}~368.07~\AA\ (dashed dotted line) and 
He~{\sc i}~584.33~\AA\ (dotted line).}
\label{fig5cds}
\end{figure}

We examined in more detail a few selected time intervals, namely A, B and C from 
the light curves, in the different spectral lines (Fig.~\ref{fig3}) as well as in the 
blue and red wings and the cores of the N~{\sc v} 1238~\AA\ and O~{\sc v} 629~\AA\ 
lines (Fig.~\ref{fig4}). During the time interval A, we noticed only a signature of 
line radiance increases/broadenings in the chromospheric and the N~{\sc v} lines 
(see their corresponding light curves in Figs.~\ref{fig3} and \ref{fig4}). No response 
was registered in O~{\sc v} or TRACE~171~\AA. In the time interval from 09:40~UT -- 09:55~UT (B) 
-- the ejection is first seen in N~{\sc v} and the chromospheric lines and 5~{\rm min} 
later in  O~{\sc v}.  The \ha\ images around 09:44~UT
show a narrow dark jet (Fig.~\ref{fig6}, upper images). The peak of the O~{\sc v} emission 
corresponds to a decrease/plateau in the N~{\sc v} and the chromospheric lines with these lines increasing again 
after O~{\sc v} starts to decrease. A  sudden rise in N~{\sc v} and the chromospheric 
lines just before 09:55~UT has no counterpart in the O~{\sc v} line.  A strong brightening in the foot-point of the 
jet precedes its occurrence. This brightening was identified as  an Ellerman bomb(s) and it 
is best seen first in \ha\ --350~m\AA\ and later also in \ha\ +350~m\AA.
It is important to mention that  a strong brightening is seen in the TRACE image at the 
same position (Fig.~\ref{fig6}). This brightening lasted and evolved for around  30~{\rm min} triggering 
a series of jets. The jets crossing the SUMER slit produced explosive events. To find 
out whether there is really a coronal response during the Ellerman bomb we studied the 
light curves of the CDS lines. The Mg~{\sc ix} 368~\AA\ line (the blend by Mg~{\sc vii} 
was removed using a double Gauss fit) shows a significant radiance increase together with 
CDS O~{\sc v}~629~\AA\  and He~{\sc i} 584~\AA\  (Fig.~\ref{fig5cds}). The changes correlated very well with 
the radiance variations of the TRACE emission from the same area. Section C 
(10:08~UT -- 10:30~UT) displays first a strong blue shifted emission in N~{\sc v} 
trailed by O~{\sc v}. Then follows a strong red shifted  
emission in the N~{\sc v} and a more modest one in O~{\sc v}. That is followed by 
another strong rise in the blue wing and the core radiance of N~{\sc v} and O~{\sc v}.   
 We found that the Ellerman bomb(s) occurs at around  10:00~UT  as a sudden 
radiance increase in all CDS spectral lines and the TRACE   (Fig.~7). At around 10:10~UT
the SUMER spectral lines (Figs.~5 and 6) show a strong dynamics as described above
 corresponding to a plasma ejection triggered by the Ellerman bomb(s).
 This complex behavior of the spectral lines during A, B and C reflects
a well know event - a surge \citep{newton}.

A surge analyzed by Tziotziou et al.(2005), employing data from the Dutch Open 
Telescope (DOT) and TRACE 1600~\AA, suggested that the temperature of the surge  
was in the region of $10^4$ to $10^5$~K, however, without access to additional data 
from higher temperatures, this was only an approximation. Here, we have data from 
O~{\sc v}~629~\AA\ and N~{\sc v}~1238~\AA, and TRACE 171~\AA\ which suggests 
heating to coronal temperatures.

We performed an identification to see how many of the brightenings seen in the N~{\sc v} 
and O~{\sc v} lines (Fig.~\ref{fig2}) can be categorized as  blinkers. The radiance 
plots in Fig.~\ref{fig4} show several radiance increases over the threshold value (1.5 
times the background emission) which can be categorized as blinkers. The blinkers, 
however, also have EEs characteristics as described above.

How can we explain the different behavior of  N~{\sc v} and O~{\sc v}?
Looking at Skylab data \citep{vern}, it can be seen that 
N~{\sc v} increases by a factor of 8 for an active region and 11.5 for a very 
active region, while the O~{\sc v} only increases by factors of 3.2 and 4.2 
respectively. As noted by Doyle et al. (2005) these lines are density 
sensitive due to density dependent ionization/recombination. For example, considering density 
dependent ionization/recombination from metastable levels we get for electron 
densities above $10^{11}$~{\rm cm}$^{-3}$ that the N~{\sc v} line can be enhanced by
almost a factor of two over that obtained using stage-to-stage 
ionization. Similar, the O~{\sc v} is reduced by almost a factor of two, 
therefore giving a factor of $\approx$3 difference in the response of N~{\sc v} 
and O~{\sc v} lines in high density plasmas. This, however, only applies to a
plasma in ionization equilibrium. For a plasma which is undergoing rapid
heating, these ratios are very different. For example, for a high-density plasma
of $10^{11}$ to $10^{12}$ cm$^{-3}$ at a temperature of logT$_e$ = 5.1  N~{\sc
v} is at 70\% of its peak while O~{\sc v} is only at 25\%  (see Fig.~4 in Doyle et al. 2005), 
hence the O~{\sc v}/N~{\sc v} ratio of the contribution functions is reduced by $\approx$3, 
very close to what is observed during the surge (see Fig.~\ref{fig4}).
More work is clearly needed in this area, for example, to address whether 
a plasma can be maintained for say several minutes at temperatures close to log T$_e$ = 5.1 in order
to have these low  O~{\sc v}~629~\AA\ to N~{\sc v}~1238~\AA\ line ratios. We present in Fig.~\ref{fig4} 
(bottom panel) the radiance ratio of O~{\sc v} versus N~{\sc v} during the surge. 

An alternative explanation is absorption due to the Lyman continuum 
\citep{kanno}, although others \citep{raymond} suggest that the effect is minimal. 
In this case, an optical thickness $\tau_H \, \approx \, 2$  is required to reduce the O~{\sc v}~629~\AA\ line 
by a factor of two. There are, however, intervals where the O~{\sc v}~629~\AA\ line is not reduced 
in intensity, which would imply a highly variable absorption, but without access to additional 
spectral lines we are unable to rule out Lyman absorption. 

\section{Conclusions}
The present study demonstrates an inter-relation between a surge observed in 
\ha\ 6563~\AA\ and SUMER explosive events registered  at transition region temperatures. 
Although some parts of the surge are seen in the TRACE 171~\AA\  images where the 
SUMER slit is positioned, the SUMER Mg~{\sc x} 625~\AA\ does not show any response. 
The surge is triggered by Ellerman bombs occurring in the region where the surge 
originates from. We found  that the Ellerman bombs reached temperatures of 1 MK from 
TRACE~171~\AA\ and CDS Mg~{\sc ix}. They are believed to be the result of magnetic 
reconnection happening in the low chromosphere triggering a surge occurrence.   With the 
present study we demonstrate that  the division of 
small-scale transient events  into a number of different subgroups, for instance explosive 
events, blinkers, spicules, surges or just brightenings, is ambiguous, implying that the 
definition of a feature based only on either spectroscopic or imaging characteristics  
as well as insufficient spectral and spatial resolution can be incomplete. 
 
In the light of present  findings, we believe that it is of great  importance to examine in detail  
the inter-relation between different transient phenomena in the solar atmosphere  using  
simultaneous Hinode and ground-based data. Hinode's Extreme-ultraviolet Imaging Spectrometer 
(EIS)  has presently the  best capabilities to provide coronal  spectral diagnostics. Equally, we 
believe that further exploration of these events by using Hinode and  the forthcoming Solar 
Dynamic Observatory data is crucial for evaluating the contribution of these  events to the 
coronal heating and slow solar wind generation. 

\acknowledgements

The authors thank the anonymous referee for the useful suggestions and comments.
The authors thank ISSI (Bern) for the support of the team ``Small-scale transient phenomena and 
contribution to coronal heating''. SUMER is part of SoHO, a collaboration between ESA and NASA 
while TRACE is a NASA SMEX mission. Research at Armagh Observatory is grant-aided
by the N.~Ireland Dept. of Culture, Arts and Leisure (DCAL). This work
was supported by STFC grants PP/E002242/1 and ST/F001843/1.


\begin{table}
\centering
\caption{The analyzed SUMER spectral lines. The expression ``/2'' means that the spectral
 line was observed in second order. The comment ``blend'' means that the spectral line is 
 blended with a close-by line. The ``a'' mark denotes lines registered by CDS.
}
\vspace{0.5cm}
\begin{tabular}{c c c c c}
\hline\hline
Ion & $\lambda/$\AA & log(T)$_{max})$/K & Comment\\
\hline

N~{\sc v} & 1238.82 & 5.3 & \\
C~{\sc i} & 1249.00 & 4.0 &  \\
O~{\sc iv}/2 & 1249.24 & 5.2&  blend\\ 
Si~{\sc x}/2 & 1249.40 & 6.1 & blend\\
C~{\sc i} & 1249.41 & 4.0 & \\     
Mg~{\sc x}/2 & 1249.90 & 6.1 & blend \\
O~{\sc iv}/2 &1250.25&5.2& blend\\
P~{\sc ii}&1249.81&4.2& blend\\
Mg~{\sc ii}&1249.93&4.1& blend\\
Si~{\sc ii} & 1250.09&4.1 &  blend\\
Si~{\sc ii} & 1250.41 &4.1 & \\
C~{\sc i} & 1250.42 & 4.0 & blend\\
S~{\sc ii} & 1250.58 &4.2 &\\
Si~{\sc ii} & 1251.16 &4.1 & \\
C~{\sc i} & 1251.17 &4.0 & blend\\
O~{\sc iv}/2&1251.70&5.2&\\
C~{\sc i} & 1252.21 & 4.0 &\\
S~{\sc ii}&1259.53&4.2&blend\\
O~{\sc v}/2 & 1259.54& 5.4&\\
O~{\sc v}$^a$&629.77&5.4&\\
Mg~{\sc vii}$^a$&368.057&5.8&blend\\
Mg~{\sc ix}$^a$&368.070&6.1&\\
He~{\sc i}$^a$&584.33&4.7\\
\hline
\end{tabular}
\label{table1}
\end{table}

\end{document}